# Spreadsheets – the Good, the Bad and the Downright Ugly

Angus Dunn
(angus@ringtailXL.com)

*Spreadsheets are ubiquitous, heavily relied on throughout vast swathes of finance, commerce, industry, academia and Government. They are also acknowledged to be extraordinarily and unacceptably prone to error. If these two points are accepted, it has to follow that their uncontrolled use has the potential to inflict considerable damage. One approach to controlling such error should be to define as "good practice" a set of characteristics that a spreadsheet must possess and as "bad practice" another set that it must avoid. Defining such characteristics should, in principle, perfectly do-able. However, being able to say with authority at a definite moment that any particular spreadsheet complies with these characteristics is very much more difficult. The author asserts that the use of automated spreadsheet development could markedly help in ensuring and demonstrating such compliance.*

## 1    GOOD AND BAD PRACTICE

It might be assumed that a good place to look for a statement of what constitutes good and bad spreadsheet practice would be EuSpRIG's own website, and more particularly, its page entitled "Best Practice". Surprisingly, though there is plenty to read there, it is hard to find best practice succinctly summarised.

Perhaps, because some have concluded that no single set of best practices are of universal application. Grossman notes that there are no "one size fits all" best practices, and that no one method is best for all situations [Thomas A Grossman, 2002] and indeed, Colver argues that EuSpRIG should not succumb to any pressure, whether internal or external, to champion any particular set of "good practices" [David Colver, 2004].

But EuSpRIG's mission was to bring together disparate groups throughout Europe "to address the ever-increasing problem of spreadsheet integrity". If it is indeed the case that no rules for creating safer spreadsheets can be compiled that are worth propagating, it is difficult to see how EuSpRIG can usefully address that problem. A very great deal has been done to help understand what is the reliability of spreadsheets and why they are less reliable than they should be. Without any statement of best practice, however, more remains to be done to help developers, be they professional or end-user, to come closer to achieving the goal of spreadsheet integrity. Any attempt to draw up a list of desirable and undesirable characteristics must have as its single objective to help these developers to achieve this goal.

But even accepting as correct the conclusion that no one size fits all, there are many common threads running through these various studies. Even David Colver acknowledges five principles as having "near universal" application. The attempt at defining best practice should seek to tease out all those near universal principles.

## 2  THE EVOLUTION OF BEST PRACTICE TO DATE

Good and bad practice has evolved in a number of strands.  Some firms which make extensive use of spreadsheets have developed their own internal sets of rules about what their spreadsheet developers must and must not do.  While such rules are probably not identical from one firm to another, they will certainly contain common elements.  Some of these firms have sought to propagate these through providing training courses to outsiders.  Academics have made studies of the prevalence of spreadsheet error and have sought to identify circumstances dangerous in the context of error and other circumstances which are regarded as safer.  Even without its own specification of good and bad practice, this is a debate to which EuSpRIG has made an important contribution.

The appendix to this paper lists a number of characteristics of spreadsheet development and suggests that each could be placed into one of four categories – required, encouraged, discouraged and forbidden.  Other practitioners and academic students of spreadsheeting might suggest that some of these characteristics did not have sufficiently universal application to justify their inclusion, would have other characteristics to add to the list and might very well argue that some of these characteristics should more appropriately belong in categories other than the ones in which they have been put.  Nevertheless, they would recognise why these characteristics had been highlighted and would understand why they have been placed where they have.  Furthermore, if a group of experienced and interested parties were to sit down with the intention of coming up with an agreed list and categorisation, they would almost certainly be successful, though possibly after some fairly spirited debate!

If "good and bad practice" is defined in this way and if this definition is promoted and becomes widely accepted and followed, an important step would have been taken towards increasing spreadsheet compliance with good practice and reducing the prevalence of spreadsheet error and thus reducing the damage that such error must inevitably cause.  For the rest of this paper this listing of good and bad practices is referred to as a code of practice.  Compliance with this code is referred to simply as "compliance".

## 3  BUT HOW TO ENFORCE IT?

Defining and promulgating good practice, however, does not ensure that such practice will be universally followed.  Even the most experienced practitioners, using all the armoury at their disposal to prevent mistakes from creeping in as they work, will make them from time to time.  Indeed some academic studies would suggest that the occurrence of errors is unavoidable [Panko 1998, revised 2008].  The requirement in some circumstances to work under huge pressure to achieve ferocious deadlines makes the probability of error even higher.  Some of these errors will be caught by the fail-safe mechanisms built in.  But some will not.

Even with the code of practice pinned to the wall in front, it will be hard, half a day into a new project, for any developer to able to say with absolute certainty that it has been complied with fully.  Since the outward appearance of a spreadsheet gives virtually no clue as to whether or not good practice has been followed, someone receiving a spreadsheet from a third party will not know at the outset whether it complies or not. Even after extended exposure to the spreadsheet, assisted by available diagnostic tools, there will still not be total certainty of compliance, short of a full high quality audit.

## 4      ONE POSSIBLE ANSWER

Various tools, such as the Operis Analysis Kit (OAK), are available to enable developers to check as they go along that they are complying with what effectively are different parts of this code of practice. If a full code of practice were adopted, there is no reason why such tools could not be extended to check compliance with every one of the points included in it. To be able to check whether a spreadsheet does or does not comply with the code at any point in time could be a useful addition to the developer's armoury. But if compliance is checked at 11:45 am and then the developer continues to develop, compliance ceases to be certified from the very moment that the developer makes the first change to a certified spreadsheet. Nevertheless, the ability to check and certify compliance at regular intervals and before the final spreadsheet is used for a critical purpose would be a marked step forward to ensuring the spreadsheet's reliability.

A spreadsheet proved compliant to a code of practice that was accepted would be a very useful first step in deriving comfort that that spreadsheet could be relied upon, since it would show the spreadsheet to be free or almost free from the structural defects which had been banned or discouraged and to possess those structural characteristics which had been encouraged or required.

Which would leave those seeking proof of its reliability free to concentrate on the appropriateness and correctness of all the logic contained in it.

## 5      ANOTHER POSSIBLE ANSWER

Compliance with codes of practice could also be a by-product of automated spreadsheet development. In *RingtailXL* [Dunn A, 2009] – full details at www.RingtailXL.com and downloadable there for evaluation - the user effectively develops a specification of the eventual Workbook. Once this has been written, the spreadsheet generator itself writes the spreadsheet in accordance with the specification the user has drafted. The spreadsheet generator can produce in parallel an easily portable and readable specification. With that specification in one hand and the spreadsheet itself in the other, the user can, if desired, read off the specification line by line and check that the spreadsheet conforms with it. It would of course be perfectly possible to develop *RingtailXL* further, so that it was capable of doing so automatically.

Much of the code of practice refers to the coding of the spreadsheet's framework and structure. Since this is hard-wired into the workings of a spreadsheet generator, when it has been proved to work once, it will work every time. After such proof, therefore, users will be able to rely on the newly generated spreadsheet as conforming with these aspects of the code of practice.

At present, *RingtailXL* generates a spreadsheet which closely follows the layout required for use within Operis, as this stood in mid-2007. Other users might have a strong preference for different layouts and structures, and, for example, for additional highlighting and formatting.

There would be nothing to stop further development to provide one or more additional spreadsheet layouts and structures, so that the user could select from those available which one was wanted.

Every spreadsheet generated by *RingtailXL* contains a macro which lights up a warning the moment any change (other than a change in data) is made to the spreadsheet. Once that warning is illuminated, the user is on notice that the spreadsheet may have ceased to comply with the code of practice.

Spreadsheets generated by *RingtailXL* allow users to change data if they wish, for example in order to use the newly generated spreadsheet to optimise the data, and thereafter to draw back into *RingtailXL* the data as changed for storage within the specification of that spreadsheet in the *RingtailXL* file. This allows the spreadsheet as it stands after the optimisation to be re-generated whenever required. Note that this feature allows changes to the data itself only; no changes to the labelling or to the structure of that data will be retained.

*RingtailXL* also facilitates compliance with certain qualitative criteria contained in the code of practice. For example, the system maintains documentation of changes made between different versions of the modules which make up the spreadsheet specification and highlights whether any such module has been changed by the user since that module was last independently checked.

## 6    CONCLUSION AND A CHALLENGE

To date, EuSpRIG has concluded that search for codified best practice would be akin to search for the end of the rainbow, pointless and to be avoided. Nevertheless, there does seem to be some degree of consensus about desirable and undesirable spreadsheet characteristics. Given that EuSpRIG exists to address the ever increasing problem of spreadsheet integrity, its aims would be furthered if it is able to help those who develop spreadsheets to make the greatest use of the essential and desirable characteristics and to avoid those that are undesirable and even downright dangerous.

A challenge, therefore, for practitioners and academics:

1. Would a code of good and bad practice be useful?

2. If so, how can compliance best be enforced and how can compliance or its absence best be recognised?

3. And finally, who would volunteer to provide comment on the appendix to this paper and input to a process which seeks an agreed code?

**ACKNOWLEDGEMENTS**:

With grateful thanks to Angela Collins for her encouragement and for her input to this paper, and to Grenville Croll for his suggestions.

**APPENDIX – SUGGESTED CODE OF GOOD AND BAD SPREADSHEET PRACTICE**

Key:   Characteristic enforced by automated spreadsheet development ✓
Characteristic encouraged/facilitated by automated spreadsheet development 👍
Characteristic possible but not yet incorporated in *RingtailXL* ✱

**REQUIRED CHARACTERISTICES**

1. All formulae cells in a range within the calculation areas of the spreadsheet to be left to right consistent ✓

2. Regions of the spreadsheet designed for data input, calculation and the presentation of output to be separated, and it should be readily apparent to the reader where these regions lie [Raffensperger 2001] ✓

3. Inputs and calculations to be modular in design [Grossman and Özlük 2004], each section representing a logically discrete portion of the total computation. The subject matter of each section should be clear, and it should be obvious to the reader as he scrolls through which section he is looking at at any time ✓

4. Related formulae should be in physical proximity; and the spreadsheet should read from left to right and from top to bottom [Raffensperger 2001]. In the calculations and inputs, related topics should where possible be recorded adjacent to each other; inputs should be recorded in the same order as are the calculations which draw on them ✓

5. If there is the slightest doubt about how the spreadsheet treats each section of the analysis, or if the spreadsheet uses one of a number of possible assumptions about how a particular item or set of items is calculated, this should be documented; the spreadsheet will generally be more readable if there is also an appropriate level of descriptive documentation as an aid to finding the way around. 👍

6. If a data book will ultimately be required, any text entries which are required to describe individual sections should be drafted at the same time as the sections themselves. 👍

7. Each section, when first drafted and after every change, should be independently checked. documentation should show whether this has been done, by whom and when. ✓

**DESIRABLE CHARACTERISTICS**

8.  A spreadsheet should be easy to read [Raffensperger 2001]. A brief glance should be enough to indicate to the reader where the inputs are, where the calculations are, where the internal checks are and which pieces contain the outputs designed to be printed out as hard copy ✓

9.  There should be a record of who drafted each section and when ✹

10. Logic once written and proven to work should be available for use within the same firm in future spreadsheets which analyse the same circumstances, until that logic is itself superseded ✓

11. The spreadsheet should be accompanied by a detailed specification. ✓

12. The hard coded parameters of the spreadsheet - start date, width, periodicity - should be input at the very end of the process, not at the very start, thus allowing these to be varied at will, without requiring changes to the template ✓

13. Multiple scenarios of inputs should be capable of being run through the same computation and there must be confidence that all inputs are brought – once and once only - into the calculations [Grossman and Özlük 2004] ✓

14. There must be extensive use of self checking ☜

15. Circularity should be avoided where possible, and controlled and tested for convergence of result where avoidance is impossible ☜

**CHARACTERISTICS TO BE AVOIDED**

16. The inclusion of hard coded constants in formulae to be discouraged ✓

17. The use of formulae of considerable complexity, rather than the use of multiple rows to achieve the same objective, each containing a part of the calculation in more transparent format X

**BANNED CHARACTERISTICS**

18. No formulae cells in a range in the calculation areas of the spreadsheet may contain a hard coded constant in place of a formula ✓

19. The default [Tools] [Options] setting for the spreadsheet should have the [Iteration] box unchecked ✓

20. The default [Tools] [Options] setting for the spreadsheet should have the [Accept labels in formulas] box unchecked ✓

21. Hidden cells and hidden sheets ✓